\documentclass[aps,
               prl,
               twocolumn,
               showpacs
              ]{revtex4}

\usepackage{amsmath, amssymb}

\usepackage{pstricks}
\usepackage{graphicx}

\renewcommand{\i}{\mathrm{i}}

\newcommand{\del}{\partial}

\newcommand{\eq}[1]{(\ref{eq:#1})}

\newcommand{\Fig}[1]{Fig.~\ref{fig:#1}}

\begin{document}

\title{Non-thermal equilibration of a one-dimensional Fermi gas}

\author{Matthias~Kronenwett}
\author{Thomas~Gasenzer}
\email{t.gasenzer@uni-heidelberg.de}
\affiliation{Institut f\"ur Theoretische Physik,
             Ruprecht-Karls-Universit\"at Heidelberg,
             Philosophenweg~16,
             D-69120~Heidelberg, Germany}
\affiliation{
             ExtreMe Matter Institute EMMI,
             GSI Helmholtzzentrum f\"ur Schwerionenforschung GmbH, 
             Planckstr.~1, 
             D-64291~Darmstadt, Germany} 

\date{\today}

\begin{abstract}

Equilibration of an isolated Fermi gas in one spatial dimension after an interaction quench is studied.
Evaluating Kadanoff-Beym dynamic equations for correlation functions obtained from the two-particle-irreducible effective action in nonperturbative approximation, the gas is seen to evolve to states characterized by thermal as well as nonthermal momentum distributions, depending on the assumed initial conditions.
For total energies near the Fermi temperature, stationary power laws emerge for the high-momentum tails while at lower momenta the distributions are of Fermi-Dirac type.
The relation found between fluctuations and dissipation exhibits nonthermal final states.
\end{abstract}


\maketitle

%
The dynamics of interacting quantum many-body systems far from equilibrium remains a challenge both for theory and experiment.
One of the seemingly simple questions concerns the long-time evolution of an isolated such system:
Given that it starts in some deliberately chosen initial non-equilibrium configuration, will the interactions between particles force the system to equilibrate, and if yes, to what kind of state?
While for classical integrable systems such problems have been studied at length since the pioneering work of Fermi, Pasta, and Ulam \cite{Ford1992a}, the evolution of quantum systems remains much more elusive.
Does a closed and finite but sufficiently large system with a ground state show, at least approximately, equilibration to a thermal or to some alternative quasi-stationary state before its unitary evolution causes revivals \cite{Greiner2002b}, and if so, on what time scale does this happen?
In recent years, new experimental techniques have been developed for ultracold atomic gases and solid-state physics that allow to precisely study quantum many-body dynamics.
Experiments have focused on the question of equilibration of a one-dimensional (1D) quantum gas with quadratic dispersion in which isolated binary collisions of the particles would not change their momenta owing to the restrictions of simultaneous momentum and energy conservation \cite{Kinoshita2006a,Hofferberth2007a}.

In this article, we study the long-time evolution and equilibration of a 1D Fermi gas containing two spin components $\alpha\in\{\uparrow,\downarrow\}$ that mutually interact through local repulsive $s$-wave collisions described by the Hamiltonian
$H=  \int d\mathrm{x}[\Psi^{\dagger}_{\alpha}(x)({-\partial_{\mathrm{x}}^{2}}/{2m})\Psi_{\alpha}(x)
  +({g_{\alpha\beta}}/{2})
    \Psi^{\dagger}_{\alpha}(x)\Psi^{\dagger}_{\beta}(x)
    \Psi_{\beta}(x)\Psi_{\alpha}(x)],
$
with $g_{\alpha\beta}=(1-\delta_{\alpha\beta})$ $4\pi a_{\mathrm{1D}}/m$, $a_{\mathrm{1D}}$ being the 1D scattering length.
$x=(x_{0}, \mathrm{x})$ includes time and space, and the complex fermionic fields obey canonical anticommutation relations
$
  \{\Psi_{\alpha}(\mathrm{x},t),\Psi^\dagger_{\beta}(\mathrm{y},t)\}
  =\delta_{\alpha\beta}\delta(\mathrm{x}-\mathrm{y})
$.
This model is integrable and has as many conserved quantities as there are degrees of freedom \cite{Yang1967a} such that it is expected not to thermalize in general, i.\,e., not to approach a grand-canonical ensemble \cite{Rigol2007a, Manmana2007a,Gangardt2008a}.
Its low-energy properties can be approximated by a Tomonaga-Luttinger liquid (TLL) \cite{Tomonaga1950a},
showing, e.\,g., a non-Fermi-Dirac power-law momentum distribution around the Fermi surface.
The long-time evolution of the TLL model, following an interaction quench, was found in \cite{Cazalilla2006a} 
to approach a generalized Gibbs ensemble \cite{Rigol2007a} determined, besides energy, by the values of conserved quasiparticle numbers.
Such non-thermal stationary states were also found in other models, both integrable and non-integrable ones
\cite{Calabrese2007a, Kollath2007a, Eckstein2008a, 
Moeckel2008a}.
 
The results of evaluating quantum-field theoretical methods we present here imply that a 1D Fermi gas defined by the full non-quadratic Hamiltonian $H$ thermalizes for sufficiently large initial energies while it can approach a non-thermal state at lower energies, with ultraviolet power-law tails in the momentum distribution away from the Fermi edge.
We consider the dynamical evolution of two-point correlation functions, including single-particle densities, as described by the Schwinger-Dyson or Kadanoff-Beym equations
\begin{align}
  &\bigl[\i
    \delta_{ac}\del_{x_{0}}
    -M_{ac}
  \bigr]
  F_{cb}(x,y)
  =\int^{x_{0}}_{t_0,z}\!
  \overline\varSigma^{\rho}_{ac}(x,z)F_{cb}(z,y)
  \nonumber\\
  &\quad  
  -\ \int^{y_{0}}_{t_0,z}\!\overline\varSigma^{F}_{ac}(x,z)\rho_{cb}(z,y),
  \nonumber\\
  &\bigl[\i
    \delta_{ac}\del_{x_{0}}
    -M_{ac}
  \bigr]
  \rho_{cb}(x,y)
  =\int^{x_{0}}_{y_{0},z}\!
 \overline\varSigma^{\rho}_{ac}(x,z)\rho_{cb}(z,y) .
  \label{eq:dynamicEqn4Frho}
\end{align}
These equations have been used to describe thermalization of both relativistic and nonrelativistic quantum gases \cite{Berges:2004yj,Gasenzer2009a}.
The statistical ($F$) and spectral ($\rho$) components of the connected two-point Green function are defined as
$G_{ab}(x,y)
     =\langle\mathcal{T} \Psi_a^\dag (x)\Psi_b(y) \rangle
  = F_{ab}(x,y) - ({\i}/{2})\rho_{ab}(x,y)$ $\mathrm{sgn}(x_{0}-y_{0})
$
where $\mathcal{T}$ denotes time-ordering,
$\mathrm{sgn}(t)$ is the sign function evaluating to $+1 (-1)$ for $t>0$ $(t<0)$.
The field index $a=(\alpha,i)$ combines the spin index $\alpha$ and the index $i$ numbering the complex fields $\Psi_{\alpha,1}(x) = \Psi_\alpha(x)$ and $\Psi_{\alpha,2}(x) = \Psi_\alpha^\dag(x)$
into a single index. 
The mean-field energy is 
$
  M_{ab}(x)
  = \delta_{ab}
     [-\partial_{\mathrm{x}}^{2}/(2m)
      - g_{\alpha\gamma}
         F_{cc}(x,x)/2]
      + g_{\alpha\beta}
         F_{ab}(x,x)
$,
where sums over spin $\gamma=\uparrow,\downarrow$ and $i_{c}=1,2$, with $c=(\gamma,i_{c})$ are implied.
The right-hand-sides of Eqs.~\eq{dynamicEqn4Frho} contain the non-Markovian memory integrals over the correlation functions 
with $\int_{t_{0},z}^{t}\equiv\int_{t_{0}}^{t}dz_{0}\int d\mathrm{z}$.
The statistical and spectral components of the ``self-energy'' 
$
  \overline\varSigma_{ab}(x,y)
  = \overline\varSigma^F_{ab}(x,y)
     - ({\i}/{2})\overline\varSigma^\rho_{ab}(x,y)$ $ \mathrm{sgn}(x_{0}-y_{0})
$
contain all the information about the correlations in the system due to interactions between particles during the past evolution.
Using a variational principle, we derive the above dynamic equations from the two-particle-irreducible (2PI) effective action $\varGamma[G]$, a functional of $G$,
see, e.\,g., Refs.~\cite{Berges:2004yj,Gasenzer2009a}.
One has $\varSigma_{ab}(x,y)=-2\i\delta\varGamma_{2}[G]/\delta G_{ba}(y,x)$ of the beyond-one-loop part $\varGamma_{2}$ of this effective action, and $\overline\varSigma_{ab}(x,y)$ contains all diagrams beyond mean-field order, $\varSigma=\varSigma^{(0)}+\overline\varSigma$, while the local part $\varSigma^{(0)}(x)$ contributes to $M$.

Different truncation schemes are available, which we found for our cases to yield qualitatively and quantitatively close results.
\begin{figure}
  \includegraphics[width=.4\textwidth]{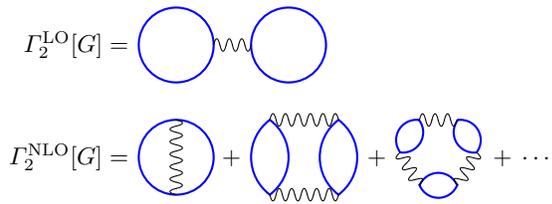}
  \caption{ (Color online) 
    Diagrams contributing in leading order (LO) and next-to-leading order (NLO) of the $1/\mathcal{N}$ expansion to $\varGamma_2$ in the two-particle-irreducible effective action.
    Bare vertices are depicted as wiggly lines at the ends of which it is summed over spin and field indices $a$.
    Vertices evaluate the incoming (blue) lines denoting $G$ at a single space-time point over which it is integrated. 
    Statistical factors are omitted.
  }
  \label{fig:Gamma2LOAndNLO}
\end{figure}
\Fig{Gamma2LOAndNLO} shows the leading-order (LO) and next-to-leading-order (NLO) contributions of the expansion in $1/\mathcal{N}$ we take into account, where $\mathcal{N}$ is the number of different spin states \cite{Berges:2001fi}.
In order to check the dependence on the truncation, we compared this with a truncation including all order-$g^{3}$ loop diagrams.

\begin{figure*}[tb]
{ \centering
  \includegraphics[width=0.97\textwidth]{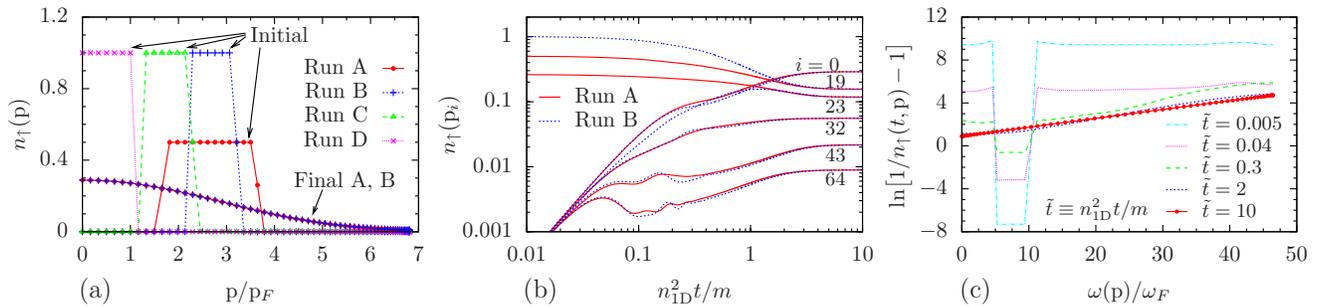}
}
  \caption{ \label{fig:RunAAndB} (color online)
    Equilibration of a one-dimensional Fermi gas starting from different initial momentum distributions.
    (a)
    Initial ($t = 0$) and final ($t=10\,mn_{\text{1D}}^{-2}$) distributions $n_\uparrow(t,|\mathrm{p}|)= n_\downarrow(t,|\mathrm{p}|)$ in runs A and B have the same total particle number and energy.
    In runs C and D, the particle number is the same while the energies are lower than in B.
    (b)
    Occupation numbers $n_\uparrow(t,|\mathrm{p}|)$ as a function of time $t$ for momentum modes $\mathrm{p}_{i}=\sin[i\pi/N_{s}]/a_{s}$,
    $N_s = 128$.
    (c)
    Inverse-slope function $\sigma(\omega_\mathrm{p})=\ln[1/n_{\uparrow}(t,\mathrm{p})-1]$ of the distribution $n_{\uparrow}(t, \mathrm{p})$ as a function of mode frequency $\omega_{\mathrm{p}}$ in units of the Fermi frequency $\omega_F$ during the thermalization process.
    $\sigma$ reduces to a straight line for a Fermi-Dirac distribution $n_{\uparrow}(\omega)=\{\exp[\beta(\omega-\mu)]+1\}^{-1}$.
  }
\end{figure*}
We solve Eqs.~\eq{dynamicEqn4Frho}, together with the integral equations determining $\overline\varSigma^{F,\rho}$, in one spatial dimension in a finite-size box with periodic boundary conditions.
The initial state is assumed to be entirely defined by the Green function $G$.
Working in Fourier space, the initial state is encoded into the $G_{ab}(x_{0},y_{0}; \mathrm{p})=\int{d} \mathrm{r}\exp[-\i \mathrm{p} \mathrm{r}] G_{ab}(x_{0},\mathrm{r};y_{0},0)$ at the initial time $x_{0}=y_{0}=t_{0}$.
Specifically, we fix the single-particle momentum distribution ($\sigma^3=$ Pauli 3-matrix)
\begin{align}
  n_{\alpha}(t,\mathrm{p}) 
  = [1 + \sigma^{3}_{ij}F_{(\alpha,j)(\alpha,i)}(t,t;\mathrm{p})]/2,
  \label{eq:SPMomDistr}
\end{align}
obtained through the definition of the statistical correlator 
  $F_{ab}(x,y)=\langle [\Psi_a^\dag(x),\Psi_b(y)] \rangle/2$,
at time $t=t_{0}$ while the spectral function
  $\rho_{ab}(x,y)=\i\langle \{\Psi_a^\dag(x),\Psi_b(y)\} \rangle$
is fixed for $x_{0}=y_{0}$ by the fermionic anticommutator, $\rho_{ab}(t,t;\mathrm{p}) = \i\delta_{ab}$.
Furthermore, we choose the coherence between different spins as well as the pair correlation function to vanish,
$\langle \Psi^{\dagger}_{\alpha}(t_0,\mathrm{x}) \Psi_{\beta}(t_0,\mathrm{x}) \rangle=0$, for $\alpha\not=\beta$, and $\langle \Psi_{\alpha}(t_0,\mathrm{x})\Psi_{\beta}(t_0,\mathrm{x}) \rangle= 0$.
This implies $F_{(\alpha,i)(\beta,j)}(t,t;\mathrm{p})= 0$, for $i\not=j$.
We restrict ourselves, here, to the special case of spin-balanced systems.

A prominent strength of the 2PI approach to far-from-equilibrium many-body quantum dynamics is that the equations of motion are derived from an action functional and therefore, for a closed system, automatically conserve total energy and further crucial quantities like the local particle current-density vector \cite{Gasenzer2009a}.
As a consequence, the dynamic equations considered here allow to study the long-time dynamics and equilibration of a system starting far away from equilibrium.

We first present results where the chosen initial conditions allow the single-particle momentum distribution, within the range of momenta considered, to thermally equilibrate.
We find that the values of the conserved quantities present in the initial state determine the final state while all other information about the details of the initial state are lost during the evolution.
\Fig{RunAAndB}a shows two different initial distributions with the same total particle number and energy from each of which the system approaches the same final state.
The coupling strength is chosen as $\gamma=mg/n_{\mathrm{1D}}=4$.
Off-energy-shell two-body scattering in the many-body background allows the momenta to be redistributed despite asymptotic conservation of momenta in isolated two-to-two scattering.
\Fig{RunAAndB}b shows the occupation numbers as a function of time $t$ for six different momentum modes.
The evolution is characterized by a short-time scale of dephasing depending on the width of the momenta present initially and a long-time scale determined by scattering.
A comparison with classical hard-sphere scattering gives that the equilibration takes on the order of 3 to 4 collision events.
We note that, for typical line densities of $n_{\mathrm{1D}}\simeq10^5\,$cm$^{-1}$, the thermalization time of a Li gas results as $\tau_{\mathrm{th}}\simeq0.25\,\mu$s.
Taking into account that the thermalization rate scales approximately as $(a_{\mathrm{1D}}n_{\mathrm{1D}})^2$ this implies thermalization times as large as seconds for a weakly interacting gas with the same $n_{\mathrm{1D}}$.

To exhibit the Fermi-Dirac character of the final momentum distribution, we show, in \Fig{RunAAndB}c, the evolution of the inverse-slope function $f(\mathrm{p})= \ln[1/n_{\uparrow}(\mathrm{p})-1]$ in run B.
For a Fermi-Dirac distribution $n_{\uparrow}(\mathrm{p})=n_\mathrm{FD}(\omega_\mathrm{p}-\mu)$, $n_\mathrm{FD}(\omega)=[\exp(\beta\omega)+1]^{-1}$,  this function is represented by a straight line with slope $\beta=1/k_{B}T$ determined by the inverse temperature and ordinate-intercept $-\beta\mu$ related to the chemical potential $\mu$. 
\begin{figure*}[tb]
{ \centering
  \includegraphics[width=0.97\textwidth]{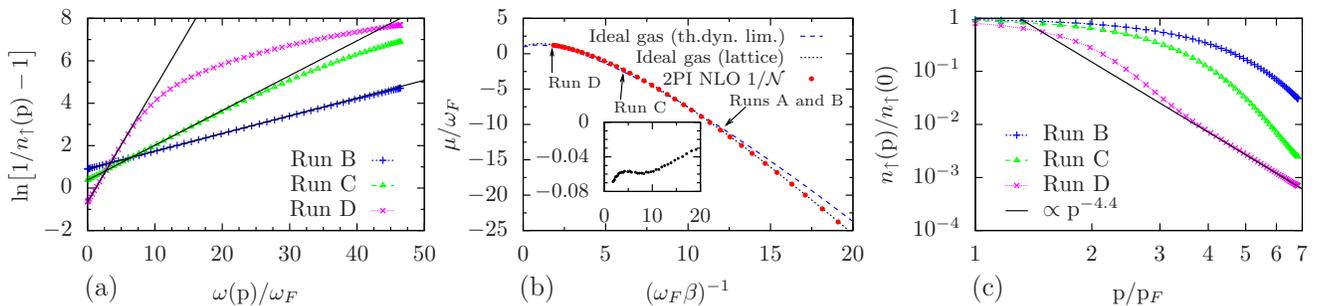}
}
  \caption{ \label{fig:RunCDAnalysis} (color online)
    (a)
    Inverse-slope functions $\sigma$ for runs B, C, and D at late times.
    The black dashed lines are Fermi-Dirac fits to the lowest 14 momentum modes from which the temperatures and chemical potentials are extracted.
    (b)
    Dependence of the chemical potential $\mu$ on the final temperature $T=(k_{B}\beta)^{-1}$, both  extracted from $\sigma$ for various runs with the same total particle number but different energies.
    (c)
    Late-time momentum distributions, equivalent to those shown in panel (a).
}
  
\end{figure*}

To investigate the character of the state approached at large times further, we compared the asymptotic momentum distributions obtained for different initial energies, for the same total particle number and interaction strength as before.
\Fig{RunAAndB}a shows the initial momentum distribution for run B and two further such distributions C and D.
Solving the equations of motion \eq{dynamicEqn4Frho}, we find that all runs reach a stationary momentum distribution; however, for lower initial energies, they do no longer correspond to a Fermi-Dirac distribution over the entire range of momenta.
As shown in \Fig{RunCDAnalysis}a, the lower momentum modes thermalize while the higher momenta in runs C and D remain overpopulated as compared to the exponential fall-off of the Fermi-Dirac distribution.
The long-time approach to these distributions is qualitatively similar to the ones shown in \Fig{RunAAndB}b, i.\,e., the distributions appear frozen on a double-logarithmic scale. 
This leads us to conjecture that they do not undergo significant further changes over a long time.
Note that the difference between the observed and thermal high-momentum occupation numbers for the respective temperatures is several orders of magnitude.

Keeping the total particle number constant,  a further reduction of the population at high momenta would not significantly alter the population of the lower momenta.
Hence, we extract temperatures and chemical potentials from a fit of the lowest 14 momentum modes to a Fermi-Dirac distribution.
The temperature dependences of the so found
chemical potential $\mu$ is shown in \Fig{RunCDAnalysis}b.
At high $T$, the chemical potential of the interacting gas ($\gamma=4$) converges to the value found for an ideal Fermi gas in thermal equilibrium on a finite-size discrete lattice, clearly distinct from the behavior in the thermodynamic limit.
At low temperatures, we find significant deviations from the ideal-gas behavior as shown in the inset.
\begin{figure}[tb]
  \includegraphics[width=0.47\textwidth]{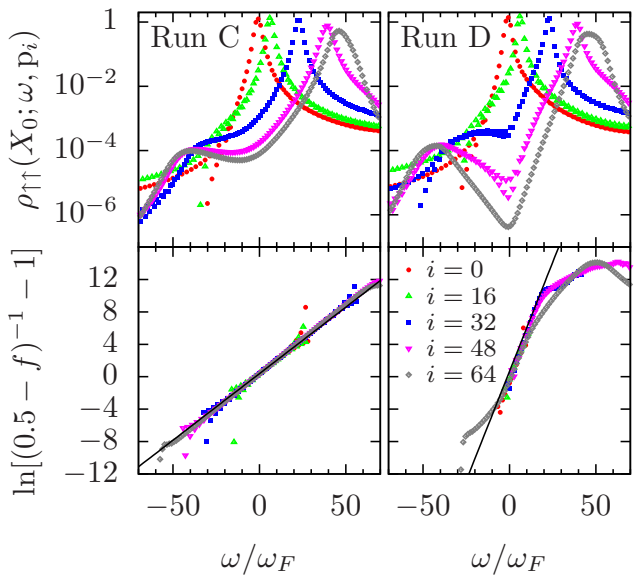}
  \caption{  \label{fig:FDT} (color online)
  Upper row: Spectral functions as a function of frequency at late time $X_{0}=18.9\,n_{\text{1D}}^{-2}m$ in runs C and D for five of the momentum modes $\mathrm{p}_{i}$, see legend in lower right panel.
  Lower row: Inverse-slope function of fractions $f$ of the statistical $F$ divided by the spectral function $\rho$ at $X_{0}=18.9\,n_{\text{1D}}^{-2}m$, for the same five momentum modes, see main text.
  Black lines indicate Fermi-Dirac distributions with $\beta$ and $\mu$ as in \Fig{RunCDAnalysis}a.
  In run D, the system does not thermalize.
}
\end{figure}

In \Fig{RunCDAnalysis}c, we depict the momentum dependence of the occupation numbers for runs C and D on a double-logarithmic scale.
This shows that the overpopulation at high momenta is characterized by a power-law $n(\mathrm{p})\propto \mathrm{p}^{-\kappa}$ with $\kappa\simeq 4.4$.
This exponent does not change when we include order-$g^3$ diagrams.
We point out that, compared to the situation in a TLL, we find algebraic fall-off at high momenta away from the Fermi edge.

This power-law tail in the momentum distribution suggests but does not automatically prove that the one-dimensional Fermi gas approaches a non-thermal state.
To investigate this point further, we have a closer look at the fluctuation-dissipation relation between the Fourier transforms of the statistical correlation function, $F_{\alpha\alpha}(X_{0};\omega,\mathrm{p})=\int ds \exp(\i\omega s)F_{\alpha\alpha}(X_{0}+s/2, X_{0}-s/2;$ $\mathrm{p})$, 
$F_{\alpha\alpha}(t,t';\mathrm{p})=\langle[\Psi^\dagger_{\alpha}(t,\mathrm{p}),\Psi_{\alpha}(t',\mathrm{p})]\rangle/2$, and the spectral function  $\rho_{\alpha\alpha}(t,t';\mathrm{p})=\i\langle\{\Psi^\dagger_{\alpha}(t,\mathrm{p}),\Psi_{\alpha}(t',\mathrm{p})\}\rangle$.
Provided that the many-body quantum state represents a grand-canonical ensemble $\rho=\exp[\beta(H-\mu N)]$ where $N$ is the total-particle-number operator, the fluctuation-dissipation theorem states that $F$ and $\rho$ are related by
\begin{equation}
  F_{\alpha\alpha}(X_{0};\omega,\mathrm{p})
   = -\i[ 1/2-n_\mathrm{FD}(\omega-\mu) ] \rho_{\alpha\alpha}(X_{0};\omega,\mathrm{p}).
\end{equation}
In \Fig{FDT}, we present, at the late time $X_{0}=18.9\,n_{\text{1D}}^{-2}m$, the fraction $f=\i F_{\uparrow\uparrow}(X_{0};\omega,\mathrm{p})/\rho_{\uparrow\uparrow}(X_{0};\omega,\mathrm{p})$ as a function of the frequency $\omega$, for five different momentum modes $\mathrm{p}_{i}$.
The lower left panel shows the inverse-slope function $\ln[(1/2-f)^{-1}-1]$ of $f$ for run C, the lower right panel for run D.
$f$ is shown in a region including the spectral peaks where the argument of the logarithm is positive.
Outside this region, it oscillates around zero, owing to the finite total evolution time after the quench.
In run C, we find, as in B (not shown), that over the region of relevant $\omega$ this function is a straight line and therefore reflects a Fermi-Dirac function.
Hence, according to the fluctuation-dissipation theorem, the system is thermalized over the depicted range of energies, despite the power-law tail in run C found for $\omega_{\mathrm{p}}\agt30\,\omega_{F}$, see \Fig{RunCDAnalysis}a.
This can be understood by considering the spectral function in \Fig{FDT} (upper row) on a logarithmic scale.
A second peak at negative frequencies, not present in an ideal gas, picks up extra contributions from the Fermi sea.
It is this second peak which causes the power-law overpopulation at high momenta, reminiscent of the BCS zero-temperature depletion of the Fermi sea in a weakly interacting system.
Hence, from run C one may preconclude that the system thermalizes to a grand-canonical ensemble, with the eigenmodes of the strongly interacting system being superpositions of particles and holes.

Run D, however, performed at even lower energy, shows that the system does in general not thermalize to a grand-canonical ensemble.
As shown in the lower right panel of \Fig{FDT}, the result of this run violates the fluctuation-dissipation theorem.
Although the momentum overpopulation is again largely produced by the contributions from the Fermi sea, see \Fig{FDT} (upper right panel), also the fraction $f$ shows a power-law tail $\sim \mathrm{p}^{-9}$.
In conclusion, we find dephasing and equilibration of a 1D Fermi gas with positive contact interactions, at sufficiently high total  energies, to a thermal state satisfying the fluctuation-dissipation theorem for a grand-canonical ensemble, while at lower total energies non-thermal stationary power-law distributions are found.

\acknowledgments \noindent 
We thank C. Bodet, M. Holland, S. Jochim, S. Kehrein, B. Nowak, J. M. Pawlowski, A. M. Rey, and D. Sexty for inspiring discussions, and JILA and the University of Colorado for their hospitality.
Support by the Deutsche Forschungsgemeinschaft, the Alliance Program of the Helmholtz Association (HA216/EMMI), the Heidelberg Graduate School for Fundamental Physics and the German Academic Exchange Service (DAAD) is acknowledged.
\\[-4.5ex]


\end{document}